\documentclass[prl,superscriptaddress,twocolumn,nofootinbib,showpacs,preprintnumbers,floatfix]{revtex4}

\usepackage{mathrsfs}
\usepackage{amsfonts,amssymb,amsmath}
\usepackage{graphicx,epsfig}
\usepackage{multirow}
\usepackage{verbatim}

\def\fsu5{$\cal{F}$-$SU(5)$}
\def\bfsu5{$\boldsymbol{\mathcal{F}}$-$\boldsymbol{SU(5)}$}

\def\m1half{$M_{1/2}$}
\def\m3half{$M_{3/2}$}
\def\m32{$M_{32}$}

\def\mt2{$M_{T2}$}
\def\x2{$\chi^2$}
\def\2b{$M_{T2}b$}

\def\bs0{$B_S^0 \rightarrow \mu^+ \mu^-$}

\def\tg{\tilde g}
\def\tq{\tilde q}
\begin{document}

\title{No Naturalness or Fine-tuning Problems from No-Scale Supergravity}

\author{Tristan Leggett}

\affiliation{George P. and Cynthia W. Mitchell Institute for Fundamental Physics and Astronomy, Texas A$\&$M University, College Station, TX 77843, USA}

\author{Tianjun Li}

\affiliation{State Key Laboratory of Theoretical Physics and Kavli Institute for Theoretical Physics China (KITPC),
Institute of Theoretical Physics, Chinese Academy of Sciences, Beijing 100190, P. R. China}

\affiliation{School of Physical Electronics, University of Electronic Science and Technology of China, 
Chengdu 610054, P. R. China }

\author{James A. Maxin}

\affiliation{Department of Physics and Engineering Physics, The University of Tulsa, Tulsa, OK 74104 USA}

\affiliation{Department of Physics and Astronomy, Ball State University, Muncie, IN 47306 USA}

\author{Dimitri V. Nanopoulos}

\affiliation{George P. and Cynthia W. Mitchell Institute for Fundamental Physics and Astronomy, Texas A$\&$M University, College Station, TX 77843, USA}

\affiliation{Astroparticle Physics Group, Houston Advanced Research Center (HARC), Mitchell Campus, Woodlands, TX 77381, USA}

\affiliation{Academy of Athens, Division of Natural Sciences, 28 Panepistimiou Avenue, Athens 10679, Greece}

\author{Joel W. Walker}

\affiliation{Department of Physics, Sam Houston State University, Huntsville, TX 77341, USA}

%%%%%%%%%%%%%%%%%%%%%%%%%%%%%%%%%%%%%%%%%%%%%%%%%%%%%%%%%%%%%%%%%%%%%%%%%%%%

\begin{abstract}
We compute the electroweak fine-tuning in No-Scale Supergravity for a representative supersymmetric
Grand Unification Theory (GUT) model, flipped $SU(5)$ with extra vector-like $flippons$, dubbed \fsu5.
We find that there is no problematic electroweak fine-tuning in No-Scale \fsu5, due to an elegant proportional rescaling of the full mass spectrum with respect to just the unified gaugino mass $M_{1/2}$, as well as a dynamic equivalence enforced between $M_{1/2}$ and the supersymmetric Higgs mixing parameter $\mu$ at the heavy unification scale.  We demonstrate both analytically and numerically that the No-Scale \fsu5 fine-tuning parameter is consequently of unit order, $\Delta_{\rm EENZ} \simeq {\cal O}(1)$, at the electroweak scale.
\end{abstract}

%%%%%%%%%%%%%%%%%%%%%%%%%%%%%%%%%%%%%%%%%%%%%%%%%%%%%%%%%%%%%%%%%%%%%%%%%%%%


\preprint{ACT-2-14}

\maketitle

%%%%%%%%%%%%%%%%%%%%%%%%%%%%%%%%%%%%%%%%%%%%%%%%%%%%%%%%%%%%%%%%%%%%%%%%%%%%

\section{Introduction}

The recent discovery at the Large Hadron Collider (LHC) of the light (126 GeV) Standard Model (SM) like
elementary Higgs boson~\cite{:2012gk,:2012gu} signals the dawn of the Beyond the SM (BSM) era. The SM is plagued by the unavoidable quadratic divergences endemic to any elementary scalar quantum field theory (QFT), which trigger the notorious destabilization known as the gauge hierarchy problem. Supersymmetry (SUSY), or boson-fermion symmetry, if broken not far above the electroweak (EW) scale, very successfully resolves the gauge hierarchy problem by transmitting ``chirality'' to the scalar member of the fermion-boson chiral supermultiplet.

Mandatory localization of this symmetry naturally induces general covariance, {\it i.e.} gravitation,
and thus also entails the spontaneous breaking of this Supergravity (SUGRA) theory via an analogous
Superhiggs mechanism.  The low energy limit (equivalently $M_{\rm Pl} \rightarrow \infty$) of spontaneously broken SUGRA consists, in its simplest form, of global SUSY with soft breaking parameters $M_{1/2},~ m_0, ~A,$ and $B$, respectively the gaugino and scalar masses, and the trilinear Yukawa and bilinear Higgs-mixing terms. Along with $\tan \beta \equiv ({v_u}/{v_d})$, the ratio of vacuum expectation values (VEVs) for 
each of the dual Higgs multiplets ($H_u$, $H_d$) required by SUSY, and the Higgs mixing term $\mu H_u H_d$, these parameters fully specify the low-energy SUSY particle spectrum, where $H_u$ and $H_d$ give masses to the up-type quarks and down-type quarks/charged leptons. Specifically, these parameters are established at some high-energy boundary scale, say $M_{\rm Unif}$, and are propagated down to physical TeV-scale masses by running of the renormalization group equations.  Generally though, SUGRA models are characterized by ($M_{\rm Pl})^4$ ``troughs'' in their effective potentials $V_{\rm eff}$, and extraneous fine-tuning remains necessary in order to achieve $V_{\rm eff} \ll \mathcal{O}(M_{\rm Pl}^4)$, even at the classical level. Furthermore, as known for more than 30 years~\cite{Ellis:1981ts}, universality of the SUSY breaking parameters at $M_{\rm Unif}$ is imperfectly mirrored at the physical EW scale, and substantial mass differentials, for example between up-squark and charm-squark, may catastrophically upset the natural suppression of flavor changing neutral currents (FCNC) in the SM!  Concurrently, a proliferation of CP-violating phases, as well as rather substantial contributions to the electric-dipole moment of the neutron (electron), escalate the game of phenomenological brinkmanship. In addition, the generic SUGRA relation $ m_0^2 \simeq M_{3/2}^2 $, where $M_{3/2}$ is the gravitino mass, implies a rather low mass gravitino $M_{3/2} \lesssim \mathcal{O}({\rm 1~TeV})$ in order to avoid fine-tuning problems at the EW scale, which stands in sharp opposition to the mass range $M_{3/2} \gtrsim \mathcal{O}({\rm 10~TeV})$ essential for avoiding cosmological tragedies~\cite{Ellis:1984eq,Ellis:1984er}. Recent LHC 8 TeV (LHC8) lower bounds on the sparticle spectrum, and moreover, the precise Planck Satellite data on cold dark matter (CDM), including inflation, do support a rather high range for the gravitino mass, thus pressing the need for a new SM. In this note we present a suitable candidate model, specifically and perhaps essentially, a No-Scale Model (NSM). Indeed, No-Scale Supergravity provides a naturally vanishing cosmological constant, at least at the classical level, and in its simplest form, a compact set of dynamically established boundary conditions $(m_0 = A = B = 0)$ that transfer the burden of SUSY breaking to the $M_{1/2}$ parameter.  In such a case, all sparticle masses acquire mass solely through gauge-mediated interactions, generating the expectation, for instance, that $m_{\widetilde{u}}^2 = m_{\widetilde{c}}^2$, and thus resolving the SUSY FCNC problem.  These boundary conditions further facilitate the rotation away of any complex CP-violating phases, resolving the $\theta_{\rm QCD}$ problem~\cite{Ellis:1982tk}. With regards to the ``gravitino problems'', $M_{1/2}$ is the real agent of mass degeneracy breaking, and hence $M_{3/2}$, which retains a role as the primordial seed of SUGRA breaking, can be safely shifted orders of magnitude above the TeV scale while still avoiding fine-tuning.  Concerning the pervasive question of fine-tuning, a quantitative measure of its value was proposed, about 30 years ago, by Ellis, Enqvist, Nanopoulos, and Zwirner (EENZ), through introduction of the parameter $\Delta$~\cite{Ellis:1986yg, Barbieri:1987fn}:
\begin{equation}
\Delta_{\rm EENZ} \equiv \left| \frac{\partial \ln (M_Z)}{\partial \ln (\widetilde{m}_i)}\right| \;,
\label{eq:delta}
\end{equation}
\noindent where $M_Z$ is the $Z$-boson (pole) mass and $\widetilde{m}_i$ generically denotes all masses
appearing in the formula defining $M_Z$ that is provided by minimization of the effective potential
$V_{\rm eff} = V_0 + V_{\rm 1-loop}$ with respect to the $H_u$ and $H_d$ directions.  Specifically:
\begin{eqnarray}
\frac{M_Z^2}{2} =
\frac{m_{H_d}^2  -
\tan^2\beta ~m_{H_u}^2}{\tan^2\beta -1} -\mu^2 \;,
\label{eq:EWMIN}
\end{eqnarray}
\noindent where $m_{H_u}^2$ and $m_{H_d}^2$ are the Higgs mass-squared values. Clearly,
large values of $\Delta _{\rm EENZ}$ will indicate large electroweak fine-tuning (EWFT): EWFT\% $\sim~ \Delta _{\rm EENZ}^{-1}$, thus large $\Delta _{\rm EENZ}$ implies small EWFT\% and large fine-tuning.

In the proposed NSM, the entire sparticle mass spectrum is driven by $M_{1/2}$, the sole non-zero soft-broken mass parameter.  Therefore, to a very good approximation, the whole sparticle spectrum rescales in linear proportion to $M_{1/2}$, in a manner reminiscent of bulk rescaling experienced by the hydrogen line-spectrum with the electron mass $m_e$! As a result, one expects from Eq. (\ref{eq:EWMIN}) for the fine-tuning to be rather small, since the unified scale $M_{1/2}$ must drop out, leaving only ratios of $\mathcal{O}$(1) coefficients. As shown below, the NSM variant dubbed No-Scale \fsu5 is further characterized by the relation
\begin{equation}
\mu \simeq M_{1/2} \;,
\label{eq:mu}
\end{equation}
\noindent which enhances the significance of the fine-tuning suppression mechanism. 
Incidentally, a relation between $\mu$ and the scalar mass $m_0$ can also reduce fine tuning~\cite{Kowalska:2014hza}.
However, the gaugino masses and trilinear soft terms may induce large fine tuning as well.
Interestingly, in our No-Scale \fsu5 model, the universal gaugino mass is the sole non-zero supersymmetry breaking soft term, therefore, we solve the SUSY electroweak fine-tuning problem completely.

\section{No-Scale \fsu5}

The No-Scale \fsu5 framework has been well detailed in the literature. The model's foundation is a tripod consisting of i) the dynamically established boundary conditions of No-Scale Supergravity, ii) the Flipped $SU(5)$ Grand Unified Theory (GUT), and iii) the derivation from local F-theory model building of a pair of hypothetical TeV-scale ``$flippon$'' vector-like super multiplets~\cite{Li:2010ws,Li:2010uu,Maxin:2011hy,Li:2011xu,Li:2011ab,Li:2013naa}.
The confluence of these attributes naturally resolves several enduring theoretical problems, while satisfying recent experimental observation~\cite{Li:2011xu,Li:2011ab,Li:2013naa}.

No-Scale Supergravity was proposed~\cite{Cremmer:1983bf} to address the cosmological flatness problem as the subspace of supergravity models to fulfill three constraints: i) the vacuum energy vanishes automatically due to the appropriate K\"ahler potential; ii) there exist flat directions that leave the gravitino mass $M_{3/2}$ undetermined at the minimum of the scalar potential; iii) the quantity ${\rm Str} {\cal M}^2$ is zero at the minimum. Large one-loop corrections would force $M_{3/2}$ to be either identically zero or of the Planck scale if the third condition were violated. A minimal K\"ahler potential that meets the first two conditions is~\cite{Ellis:1984bm,Cremmer:1983bf}
\begin{eqnarray} 
K &=& -3 {\rm ln}( T+\overline{T}-\sum_i \overline{\Phi}_i
\Phi_i)~,~
\label{NS-Kahler}
\end{eqnarray}
where $T$ is a modulus field and $\Phi_i$ are matter fields, which parameterize the non-compact $SU(N,1)/SU(N) \times U(1)$ coset space. The third condition can always be satisfied in principle and is model dependent~\cite{Ferrara:1994kg}. From the K\"ahler potential in Eq.~(\ref{NS-Kahler}) one automatically obtains the No-Scale boundary conditions $m_0 = A = B = 0$, while $M_{1/2}$ can be non-zero and evolve naturally, as is in fact required for SUSY breaking. The high-energy boundary condition $B = 0$ effectively fixes $\tan\beta$ at low energy. The gravitino mass $M_{3/2}$ is determined by the equation $d(V_{EW})_{min}/dM_{3/2}=0$, since the minimum of the electroweak (EW) Higgs potential $(V_{EW})_{min}$ depends on $M_{3/2}$, and the supersymmetry breaking scale is thus determined dynamically. The result is a natural $one$-$parameter$ model, with $M_{1/2}$ the single degree of freedom.

Exact string-scale gauge coupling unification, while also avoiding the Landau pole problem, can be
accomplished by supplementing the standard ${\cal F}$-lipped $SU(5)\times U(1)_X$~\cite{Nanopoulos:2002qk,Barr:1981qv,Derendinger:1983aj,Antoniadis:1987dx} SUSY field content with the following TeV-scale vector-like multiplets
($flippons$)~\cite{Jiang:2006hf}
\begin{eqnarray}
\hspace{-.3in}
& \left( {XF}_{\mathbf{(10,1)}} \equiv (XQ,XD^c,XN^c),~{\overline{XF}}_{\mathbf{({\overline{10}},-1)}} \right)\, ,&
\nonumber \\
\hspace{-.3in}
& \left( {Xl}_{\mathbf{(1, -5)}},~{\overline{Xl}}_{\mathbf{(1, 5)}}\equiv XE^c \right)\, ,&
\label{z1z2}
\end{eqnarray}
where $XQ$, $XD^c$, $XE^c$, $XN^c$ have the same quantum numbers as the quark doublet, the right-handed down-type quark, charged lepton, and neutrino, respectively. Models of this nature can be achieved in ${\cal F}$-ree ${\cal F}$-ermionic string constructions~\cite{Lopez:1992kg} and ${\cal F}$-theory model
building~\cite{Jiang:2009zza,Jiang:2009za}, and are referred to as \fsu5~\cite{Jiang:2009zza}.

A minimal set of requisite constraints from theory and phenomenology~\cite{Li:2011xu,Li:2013naa} is satisfied by the No-Scale \fsu5 model space. The set of minimal constraints satisfied are as follows: i)
consistency with the dynamically established boundary conditions of No-Scale SUGRA (most importantly the strict imposition of a vanishing $B$ parameter at the ultimate \fsu5 unification scale $M_{\cal F}$ near $M_{\rm Pl}$, enforced as $\left|B\right(M_{\cal F})| \leq 1$ GeV, commensurate with the scale of EW radiative corrections); ii) radiative electroweak symmetry breaking (REWSB); iii) the measured central value of the Planck CDM relic density $\Omega h^2 = 0.1199 \pm 0.0027$~\cite{Ade:2013zuv}); iv) the world average top-quark mass $m_t = 173.3 \pm 1.1$~GeV~\cite{:1900yx}; v) precision LEP constraints on the light SUSY chargino and neutralino mass content~\cite{LEP}; and vi) production of a lightest CP-even Higgs boson mass of $m_{h} = 125.5 \pm 1.5$ GeV.  With regards to the last criterion, it has been observed that additional tree level and one-loop contributions to the Higgs boson mass are supplied by interaction with the $flippon$ supermultiplets~\cite{Li:2011ab,Li:2013naa}, which may augment the Minimal Supersymmetric Standard Model (MSSM) prediction by as much as 3-5 GeV, in a manner that is more effective for lighter values of the flippon mass $M_V$.

A two-dimensional parameterization in the vector-like $flippon$ super-multiplet mass scale $M_V$ and
the universal gaugino boundary mass scale $M_{1/2}$ is extracted via application of the prior constraints
from a larger four-dimensional hyper-volume that also comprises the top quark mass $m_t$ and the ratio $\tan \beta$. The surviving model space consists of a diagonal wedge ({\it cf.} Ref.~\cite{Li:2013naa}) in ($M_{1/2}$, $M_V$) space, the extent of which is bounded at small $M_{1/2}$ and small $M_V$ by the LEP constraints, and at large $M_{1/2}$ and large $M_V$ by the CDM constraints and transformation to a charged stau LSP. Likewise, the upper limit at larger $M_V$ and lower limit at smaller $M_V$ are constrained by the central experimental range on the top quark mass. The intersection of all constraints nets an experimentally viable model space ranging from $M_{1/2} \simeq 400$ GeV to $M_{1/2} \simeq 1500$ GeV, with a corresponding vector-like $flippon$ mass of $M_V \simeq 1$ TeV to $M_V \simeq 180$ TeV and $\tan\beta \simeq 19.5$ to $\tan\beta \simeq 25$. Our results suggest that the lower scale bound in No-Scale \fsu5 derived from LHC SUSY searches coincides with a gaugino mass in the vicinity of $M_{1/2}$ = 850--1000 GeV, which corresponds to a gluino mass of $M_{\tg}$ = 1150--1300 GeV.  This independent data-driven
demarcation is broadly compatible with the limits on simplified model scenarios published by the ATLAS
and CMS Collaborations for the LHC8 run, which indicate lower gluino mass constraints of about 1.7 TeV and 1.3 TeV for $m_{\tg} \sim m_{\tq}$ and $m_{\tg}\ll m_{\tq}$, respectively~\cite{ATLAS,CMS}.

Furthermore, a recent analysis~\cite{Ellis:2013xoa,Ellis:2013nxa,Ellis:2013nka} of the Planck satellite measurements suggests a mechanism, given appropriate superpotential parameter choices, for mimicking the Starobinsky model of inflation~\cite{Starobinsky:1980te,Mukhanov:1981xt,Starobinsky:1983zz}, and thus naturally enforcing compatibility between the data and cosmological models based upon No-Scale SUGRA.
The Starobinsky model represents an ad-hoc adaptation of Einstein's description of gravity, which combines
a quadratic power of the Ricci scalar with the standard linear term. There is considerable enthusiasm for the realization that this esoteric $(R+R^2)$ model is conformally equivalent to No-Scale supergravity with an $SU(2,1)/SU(2) \times U(1)$ K\"ahler potential~\cite{Ellis:2013xoa,Ellis:2013nxa,Ellis:2013nka}, which is a subcase of Eq.~(\ref{NS-Kahler}).  Specifically, the algebraic equations of motion corresponding to an
auxiliary scalar field $\Phi$ with a quadratic potential that couples to a conventional Einstein term may be substituted back into the action, resulting in the quadratic power of the scalar curvature~\cite{Stelle:1977ry,Whitt:1984pd} that is phenomenologically favorable.

\section{ A Natural Solution to the Supersymmetric Electroweak Fine-Tuning Problem}

In the MSSM, the minimization condition from the Higgs scalar potential, which determines the $Z$-boson mass, is given in Eq.~(\ref{eq:EWMIN}). For a moderately large $\tan\beta$, we get
\begin{eqnarray}
\frac{M_Z^2}{2} \simeq -m_{H_u}^2 -\mu^2 \;.
\label{eq:EWMINL}
\end{eqnarray}
\noindent As such, if both $-m_{H_u}^2$ and $\mu^2$ are large, it is necessary to fine-tune both parameters to realize the correct $Z$-boson mass from large cancellations. In the CMSSM/mSUGRA, $\mu$ is determined by the electroweak symmetry breaking condition, and a small $-m_{H_u}^2$ is thus required to evade the electroweak fine-tuning problem.  However, the stop quarks and gluino will problematically
provide dominant contributions to $-m_{H_u}^2$ at one and two loops respectively. Additionally, in order to lift the Higgs boson mass to about 125.9 GeV radiatively, the MSSM stop quarks must likewise be quite heavy, directly threatening to reanimate the very gauge hierarchy problem that SUSY was originally intended to solve.

In gravity mediated supersymmetry breaking, the $\mu$ problem, which pertains to the Higgs bilinear mass $\mu$ term near the electroweak scale, can in fact be solved via the Giudice-Masiero (GM) mechanism~\cite{Giudice:1988yz}. Applying the GM mechanism in No-Scale SUGRA gives $\mu \propto M_{1/2} \propto M_{3/2}$, suggesting mutual proportionality, where the gravitino mass can be situated around 30 TeV to elude the gravitino problem. Crucially, finding $\mu \simeq M_{1/2}$ in No-Scale \fsu5 approximately rescales the sparticle spectra per variation only in $M_{1/2}$.  Possessing such a property, we can show that there is no residual electroweak fine-tuning problem, as follows.

Given $\mu \simeq M_{1/2}$, we shall fix the \fsu5 unification scale $M_{\cal F}$, and also $\tan\beta$,
the top quark mass $m_t$, top quark Yukawa coupling, and all remaining input parameters at the unification scale, with the exception of the vector-like $flippon$ mass $M_V$. However, compensation by $M_V$ is quite weak, being transmitted by logarithmic threshold corrections. As a result, $M_{Z}$ is a trivial function of $M_{1/2}$, 
and we have the following approximate scale relation
\begin{eqnarray}
M_Z^n ~=~ f_n \left( c_i \right) ~M_{1/2}^n~,~\,
\label{eq:fn}
\end{eqnarray}
where $f_n$ is a dimensionless parameter and approximately
a constant in No-Scale \fsu5 due to the rescaling property in terms of $M_{1/2}$. Also,
$c_i$ denote the dimensionless parameters, 
for example, the gauge and Yukawa couplings, as well as the ratios between $\mu$ and $M_{1/2}$
and between $M_V$ and $M_{1/2}$, etc.
Although we can take $n=2$, to make it generic, we do not fix an integer number for $n$.
In addition, one naive question is what the low energy experimental input $M_Z$ 
determines in our model due to the rescaling property. Our later study showed 
that it determines the ratio between $\mu$ and $M_{1/2}$~\cite{Leggett:2014hha}.

In gravity mediated supersymmetry breaking, the typical quantitative measure $\Delta_{\rm EENZ}$ for fine-tuning is the maximum of the logarithmic derivative of $M_Z$ with respect to all fundamental parameters $a_i$ at the GUT scale~\cite{Ellis:1986yg, Barbieri:1987fn}
\begin{eqnarray}
\Delta_{\rm EENZ} ~=~ {\rm Max}\{\Delta_i^{\rm GUT}\}~,~~~
\Delta_i^{\rm GUT}~=~\left|
\frac{\partial{\rm ln}(M_Z)}{\partial {\rm ln}(a_i^{\rm GUT})}
\right|~.~\,
\label{eq:BG-EENZ}
\end{eqnarray}

\noindent For the nearly constant $f_n$ of Eq. (\ref{eq:fn}) in No-Scale \fsu5, we have
\begin{eqnarray}
\frac{\partial M_Z^n}{\partial M_{1/2}^n} ~\simeq~ f_n~,~\, 
\label{eq:fnd}
\end{eqnarray}
\noindent and therefore we obtain
\begin{eqnarray}
\frac{\partial{\rm ln}(M_Z^n)}{\partial {\rm ln}(M_{1/2}^n)}
~\simeq~  \frac{M_{1/2}^n}{M_Z^n} \frac{\partial M_Z^n}{\partial M_{1/2}^n} ~\simeq~ \frac{1}{f_n} f_n~.~\
\label{eq:fnpar}
\end{eqnarray}

\noindent Consequently, the fine-tuning is an order one constant,
\begin{eqnarray}
\left|
\frac{\partial{\rm ln}(M_Z^n)}{\partial {\rm ln}(M_{1/2}^n)}
\right| \simeq {\cal O}(1)~.~\,
\label{eq:orderone}
\end{eqnarray}

\noindent Therefore, there is no electroweak fine-tuning problem in No-Scale \fsu5. We shall confirm this here via numerical calculations of $\Delta_{\rm EENZ}$ for $n = 2$ in Eq.~(\ref{eq:orderone}).

\section{Calculation of $\Delta_{\rm EENZ}$ and EWFT}

\begin{figure}[htp]
        \centering
        \includegraphics[width=0.5\textwidth]{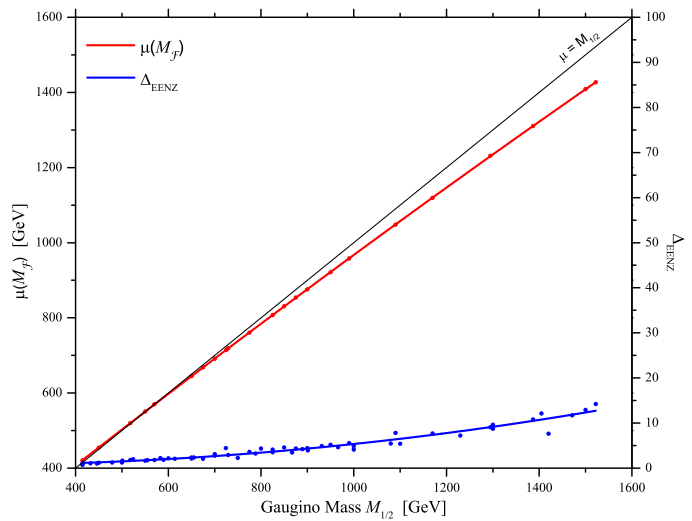}
        \caption{Depiction of the $\mu \simeq M_{1/2}$ relationship in the No-Scale \fsu5 model space (top curve -- left scale), with the $\mu$ parameter computed at the unification scale $M_{\cal F}$. The $\Delta_{\rm EENZ}$ parameter is also shown (bottom curve -- right scale), with both curves a function of $M_{1/2}$.}
        \label{fig:delta}
\end{figure}

The $Z$-boson mass is typically an input into a Renormalization Group Equation (RGE) network iteratively evolved from the unification scale down to the electroweak scale. Therefore, in order to verify Eq. (\ref{eq:orderone}) for $n=2$, we numerically fluctuate Eq. (\ref{eq:EWMIN}), using a proprietary modification of the {\tt SuSpect~2.34}~\cite{Djouadi:2002ze} codebase to run $flippon$ enhanced RGEs.  Holding $M_{\cal F}, ~\tan\beta, ~m_t,$ etc. constant, we take the partial derivative of Eq. (\ref{eq:EWMIN}) with respect to the sole floating parameter $M_{1/2}^2$, giving
\begin{eqnarray}
\frac{1}{2}\frac{\partial M_Z^2}{\partial M_{1/2}^2} = \frac{\frac{\partial m_{H_d}^2}{\partial M_{1/2}^2}   - \tan^2\beta ~\frac{\partial m_{H_u}^2}{\partial M_{1/2}^2}}{\tan^2\beta -1} -\frac{\partial \mu^2}{\partial M_{1/2}^2} \;.
\label{eq:EWMINd}
\end{eqnarray}
\noindent Permission to fix $\tan\beta$ in Eq. (\ref{eq:EWMINd}) is granted by transiting along a contour of constant $\tan\beta$~\cite{Li:2013naa} from the benchmark point, necessitating a compensating adjustment in the vector-like $flippon$ mass $M_V$, though as described, the transmission of this $M_V$ effect is weak, and hence negligible. Next, we compute the three partial derivatives of $M_{H_d}^2, ~M_{H_u}^2,$ and $\mu^2$ with respect to $M_{1/2}^2$ at the scale of electroweak symmetry breaking, denoted as $Q_0$. To accomplish this, we vary $M_{1/2}$ locally for a set of benchmark points of the No-Scale \fsu5 model space~\cite{Li:2013naa}, fixing all the aforementioned parameters. We compute $M_{H_d}^2, ~M_{H_u}^2,$ and $\mu^2$ at $Q_0$ via the RGEs, then determine the local slope around each benchmark
point of the $(M_{1/2}^2,M_{H_d}^2)$, $(M_{1/2}^2,M_{H_u}^2)$, and $(M_{1/2}^2,\mu^2)$ diagrams to numerically calculate the derivatives needed to compute Eq. (\ref{eq:EWMINd}). We then use Eq. (\ref{eq:fnpar}) with $n=2$ to find $\Delta_{\rm EENZ}$, as illustrated in the lower curve of Fig. (\ref{fig:delta}). The top curve of Fig. (\ref{fig:delta}) clearly exhibits the relation of Eq. (\ref{eq:mu}), displaying the near exact intrinsic correlation between the SUSY mass parameter $\mu$ at the No-Scale \fsu5 unification scale $M_{\cal F}$ and the gaugino mass $M_{1/2}$, also defined at $M_{\cal F}$.

Observe in Fig. (\ref{fig:delta}) how the $\Delta_{\rm EENZ}$ parameter tracks the delta between
$\mu(M_{\cal F})$ and $M_{1/2}$. As $\mu$ and $M_{1/2}$ begin to slightly diverge at large $M_{1/2}$,
even just negligibly, the sensitivity of the $\Delta_{\rm EENZ}$ parameter to this behavior is highly
precise. Likewise, at small $M_{1/2}$, comparable action is witnessed where $\mu$ and $M_{1/2}$ are
explicitly equal and thus $\Delta_{\rm EENZ}$ is practically zero. This does indeed verify the ${\cal O}(1)$
result of Eq. (\ref{eq:orderone}) for the No-Scale \fsu5 model space, indicating no electroweak
fine-tuning as a result of the rescaling property in terms of $M_{1/2}$ and inherent correspondence of
$\mu \simeq M_{1/2}$.

\section{Conclusions}

The No-Scale model framework, and in particular the phenomenologically viable representative No-Scale \fsu5, effectively confronts the problems of naturalness in the MSSM, thanks to compensating action of the boundary conditions ($m_0 = A = B = 0$), global spectral dependence on the isolated soft-breaking SUSY parameter $M_{1/2}$, and the $\mu \simeq M_{1/2}$ relation, implying:

\medskip \noindent i) No FCNC problem. \\
ii) No CP-violating phenomenology problem. \\
iii) No cosmological gravitino problem. \\
iv) Natural No-Scale realization of the Starobinsky inflationary model, where the inflaton is
identified with the sneutrino, leading to sufficient reheating and baryon-asymmetry. \\
v) No LHC8 observational problem. \\
vi) No CDM problem. \\
vii) No EWFT problem as expressed by $\Delta_{\rm EENZ} \simeq {\cal O}(1)$. \\

%%%%%%%%%%%%%%%%%%%%%%%%%%%%%%%%%%%%%%%%%%%%%%%%%%%%%%%%%%%%%%%%%%%%%%%%%%%%

\begin{acknowledgments}
This research was supported in part by the DOE grant DE-FG03-95-Er-40917 (DVN), by the Natural Science
Foundation of China under grant numbers 10821504, 11075194, 11135003, 11275246, and 11475238,
and by the National Basic Research Program of China (973 Program) under grant number 2010CB833000 (TL).
We also thank Sam Houston State University for providing high performance computing resources.
\end{acknowledgments}

%%%%%%%%%%%%%%%%%%%%%%%%%%%%%%%%%%%%%%%%%%%%%%%%%%%%%%%%%%%%%%%%%%%%%%%%%%%%

\bibliography{bibliography}

\end{document}